\documentclass[%
pra,
twocolumn,
amsmath,
amssymb,
reprint,
superscriptaddress,
groupedaddress,
footinbib,
 amsmath,amssymb,
 aps,
longbibliography
]{revtex4-2}
\usepackage[colorlinks=true,citecolor=blue,linkcolor=black]{hyperref}
\usepackage{lipsum}
\usepackage{float}
\usepackage{graphicx}
\usepackage[utf8]{inputenc}
\usepackage[english]{babel}
\usepackage{amsmath,amsfonts,amssymb}
\usepackage[T1]{fontenc}
\usepackage{url}
\usepackage{changes}
\usepackage{amsmath}
\usepackage{amsfonts}
\usepackage{amssymb}
\usepackage[version=3]{mhchem}
\usepackage{changes}
\definechangesauthor[color=red]{TJK}
\usepackage{changes}
\usepackage{soul}

\begin{document}

\title{Coherent terahertz-to-microwave link using electro-optic-modulated Turing rolls}
\author{Wenle Weng}
\email[]{wenle.weng@epfl.ch}
\affiliation{Institute of Physics, Swiss Federal Institute of Technology (EPFL), CH-1015 Lausanne, Switzerland}

\author{Miles H. Anderson}
\affiliation{Institute of Physics, Swiss Federal Institute of Technology (EPFL), CH-1015 Lausanne, Switzerland}

\author{Anat Siddharth}
\affiliation{Institute of Physics, Swiss Federal Institute of Technology (EPFL), CH-1015 Lausanne, Switzerland}

\author{Jijun He}
\affiliation{Institute of Physics, Swiss Federal Institute of Technology (EPFL), CH-1015 Lausanne, Switzerland}

\author{Arslan S. Raja}
\affiliation{Institute of Physics, Swiss Federal Institute of Technology (EPFL), CH-1015 Lausanne, Switzerland}

\author{Tobias J. Kippenberg}
\email[]{tobias.kippenberg@epfl.ch}

\affiliation{Institute of Physics, Swiss Federal Institute of Technology (EPFL), CH-1015 Lausanne, Switzerland}

\begin{abstract}
Arising from modulation instability, Turing rolls in optical Kerr microresonators have been used in the generation of optical frequency combs and the synthesis of microwave and terahertz frequencies. In this work, by applying electro-optic modulation on terahertz-frequency Turing rolls, we implement electro-optic frequency division with a microcomb to synthesize variable low-noise microwave signals. We also actively stabilize the terahertz oscillations to a microwave reference via intracavity power modulation, obtaining fractional frequency instabilities that are better than those of the free-running situation by up to six orders of magnitude. This study not only highlights the extraordinary spectral purity of Turing roll oscillations but also opens the way for bidirectional terahertz-to-microwave links with hybrid optical frequency comb techniques.
\end{abstract}

\maketitle
\section{Introduction}

As ``one of the most ubiquitous types of instabilities in nature'' \cite{zakharov2009modulation}, modulation instability (MI) has been found in a variety of physical systems ranging from hydrodynamics \cite{benjamin1967disintegration}, plasma \cite{thornhill1978langmuir}, optics \cite{tai1986observation,kip2000modulation}, to ultracold gases \cite{strecker2002formation,khaykovich2002formation,konotop2002modulational}. In optical Kerr microresonators pumped by continuous-wave (cw) lasers, MI gives rise to cavity-enhanced optical frequency generation through four-wave mixing \cite{godey2014stability}, leading to the discovery of Kerr microcombs \cite{del2007optical,kippenberg2011microresonator}. Although in recent years the research focus of the microcomb community has shifted to the state of dissipative Kerr solitons (DKS) \cite{herr2014temporal,kippenberg2018dissipative} for its exceptional dynamical stability and the intrinsic broad spectral span, the state of MI-induced Turing rolls (TRs) is easier to access and can be advantageous in certain applications such as efficient terahertz (THz) excitation \cite{huang2017globally}, coupled laser-microresonator comb generation \cite{bao2020turing}, perfect soliton crystal inception \cite{karpov2019dynamics}, and noise-robust fiber communications \cite{pfeifle2015optimally}. Also known as cnoidal waves, TRs are periodic pulse-like solutions to the Lugiato-Lefever equation (LLE) -- the master equation that describes the cw-driven Kerr cavity system \cite{lugiato1987spatial,haelterman1992dissipative}. Aside from the many applications in optics and photonics, TRs have been attracting interests for their fundamental complex properties and particularly their close relation to other nonlinear dynamical states \cite{copie2016competing,qi2019dissipative,kholmyansky2019optimal}.

Owing to the compactness and the simplicity of the setup, microwave generation based on TRs in crystalline microresonators have been demonstrated \cite{liang2015high,saleh2016phase}. Yet, compared with the DKS counterparts \cite{liang2015high,weng2019spectral,lucas2020ultralow}, TR oscillations utilized in prior studies exhibit substantially higher phase noise. Moreover, in a microresonator with an ultrahigh quality factor ($Q$), the frequency of the TR oscillation is usually a high multiple of the cavity free spectral range (FSR). As a result, the oscillation frequency can fall into the THz range that cannot be detected and measured with conventional electronics. Although the photodetection and the optical-to-THz signal conversion of THz-repetition-rate microcombs have been realized with advanced electronic-photonic solutions \cite{huang2017globally,zhang2019terahertz}, they haven't reached the technical maturity that is required for wide-spread applications. Consequently, TRs' usage in low-noise signal synthesis and metrology at radiofrequency (rf) and THz ranges is severely hindered.

In this work, we overcome the twofold disadvantage of TRs by adopting the electro-optic frequency division (EOFD) technique \cite{li2014electro} to synthesize low-noise microwaves from a THz TR oscillation. The developed TR-EOFD scheme allows the spectral quality of a TR oscillation to be faithfully transferred into variable microwave frequencies with the benefit of a substantial reduction of phase noise due to the frequency division mechanism. With the synthesized 10-GHz signal reaching phase noise levels of $-60$ and $-110$ dBc$/$Hz at offset frequencies of 10 Hz and 1 kHz respectively, the spectral purity of the microwaves is comparable to the best performances ever achieved with DKS \cite{liang2015high,lucas2020ultralow}. In addition, the system can be easily adapted to transfer the frequency stability of an rf reference to the TRs, thus enabling a bidirectional coherent THz-to-microwave link that could potentially find a variety of applications in sensing, telecommunication and metrology.

\begin{figure*}[hbt]
\centering
\includegraphics[width=1.76\columnwidth]{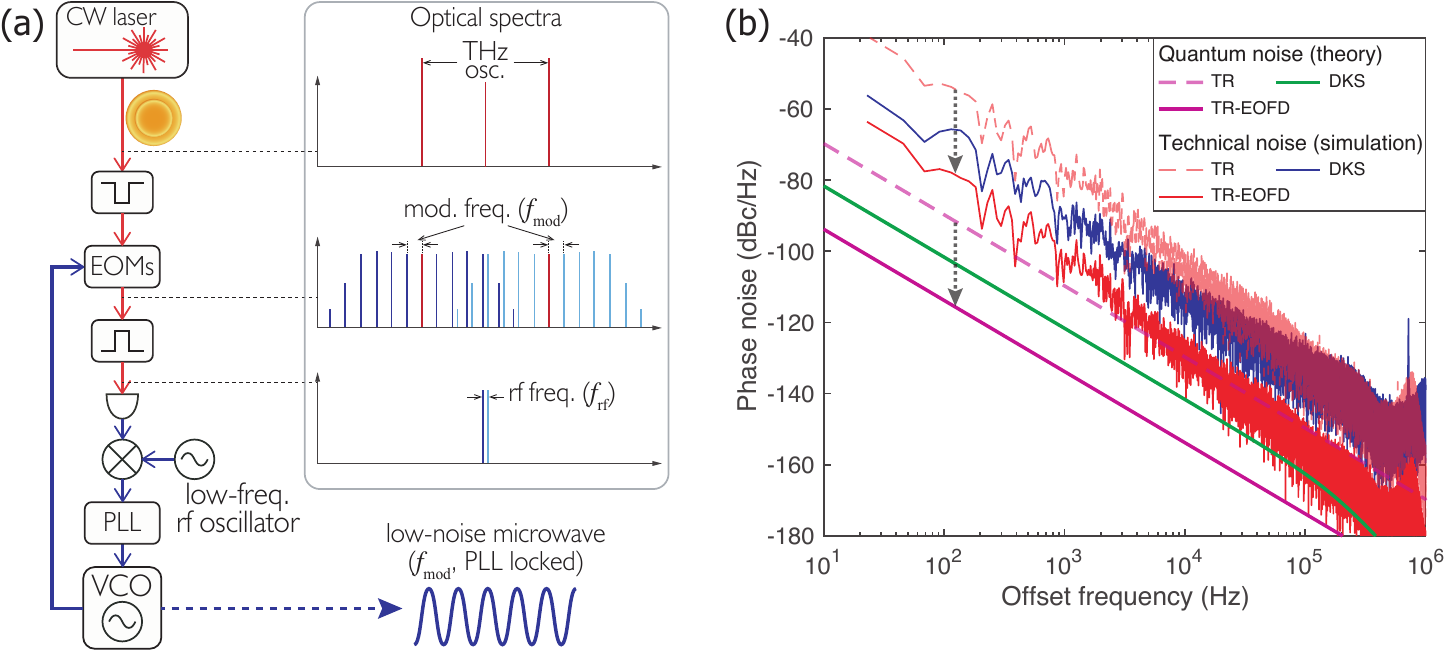} 
\caption{Low-noise microwave synthesis with the TR-EOFD technique. (a) The schematic layout of the TR-EOFD setup. (b) The phase noise spectra of computed fundamental quantum noises and simulated technical noises of a terahertz TR oscillation, a single-soliton DKS oscillation, and a microwave synthesized with the TR-EOFD scheme, respectively. The  synthesized microwave has a frequency that is the same as the DKS repetition rate (12~GHz in the simulations, details of which are presented in Appendix A and Appendix B). The black arrows indicate the phase noise reduction due to the frequency division.
}\label{fig1}
\end{figure*}

\section{Principle}

Figure \ref{fig1} (a) shows the principle of the TR-EOFD scheme. A cw-laser-pumped microresonator produces a TR oscillation with a frequency in the THz range, corresponding to two intensive first-order TR sidebands (and multiple higher-order sidebands with much weaker amplitudes) in the optical spectrum. After proper filtering, the two TR sidebands pass through cascaded electro-optic phase modulators (EOMs) that are modulated with a microwave frequency $f_{\rm mod}$ derived from a voltage-controlled oscillator (VCO). If the modulation is strong enough, each TR sideband becomes an electro-optic frequency comb (EOC) \cite{del2012hybrid,parriaux2020electro,diddams2020optical} whose spectral width can cover the frequency of the THz TR oscillation ($f_{\rm TR}$). Two closely separated comb teeth, each of which is generated from a different first-order TR sideband, can be filtered out at the center of the two TR sidebands to produce a beat frequency $f_{\rm rf}$. When the VCO is free running, the single-sideband phase noise of the beat frequency signal $\mathcal{L}_{\rm rf}$ can be expressed as:
\begin{equation}
\label{L_free}
\mathcal{L}_{\rm rf} = \mathcal{L}_{\rm TR} + n^2 \mathcal{L}_{\rm VCO}
\end{equation}
where $\mathcal{L}_{\rm TR}$ is the phase noise of the THz TR oscillation. The second term on the right-hand side represents the additional phase noise caused by the multiplication of the free-running VCO noise in EOC \cite{parriaux2020electro,diddams2020optical}, and the integer $n \approx \frac{f_{\rm TR}}{f_{\rm mod}}$ is the division factor. By tuning $f_{\rm mod}$, $f_{\rm rf}$ can be tuned to closely match the frequency of a reference rf oscillator (e.\,g., a 10-MHz reference signal). With a phase-locked loop (PLL), the VCO can be controlled to lock $f_{\rm rf}$ to the rf reference, and then $f_{\rm mod}$ is related to $f_{\rm TR}$ as:
\begin{equation}
\label{principle}
f_{\rm mod} = \frac{f_{\rm TR} + (-) f_{\rm rf}}{n} 
\end{equation}
where the sign in the second term depends on the relative positions of the two teeth. One should note that in principle the two teeth can be set to have the exactly same frequency in order to eliminate the need for the reference rf oscillator. In practice, however, using a beat frequency in the rf domain for the servo facilitates the locking because the error signal can be conveniently amplified by a rf amplifier without being interfered by the relatively high technical noises (e.\,g., electronic noise in the photodetector and laser intensity fluctuations) in the low frequency range. Because normally the phase noise of the rf reference signal is extremely low, within the PLL locking bandwidth the phase noise of the controlled VCO is:
\begin{equation}
\label{division}
\mathcal{L}'_{\rm VCO} = \frac{\mathcal{L}_{\rm TR}}{n^2}
\end{equation}
showing that the same noise characteristics of the THz TR oscillation is inherited by the controlled VCO, with an additional reduction by a factor of $\frac{1}{n^2}$ in the phase noise magnitude \cite{li2014electro}.

\begin{figure*}[hpt]
\centering
\includegraphics[width=2\columnwidth]{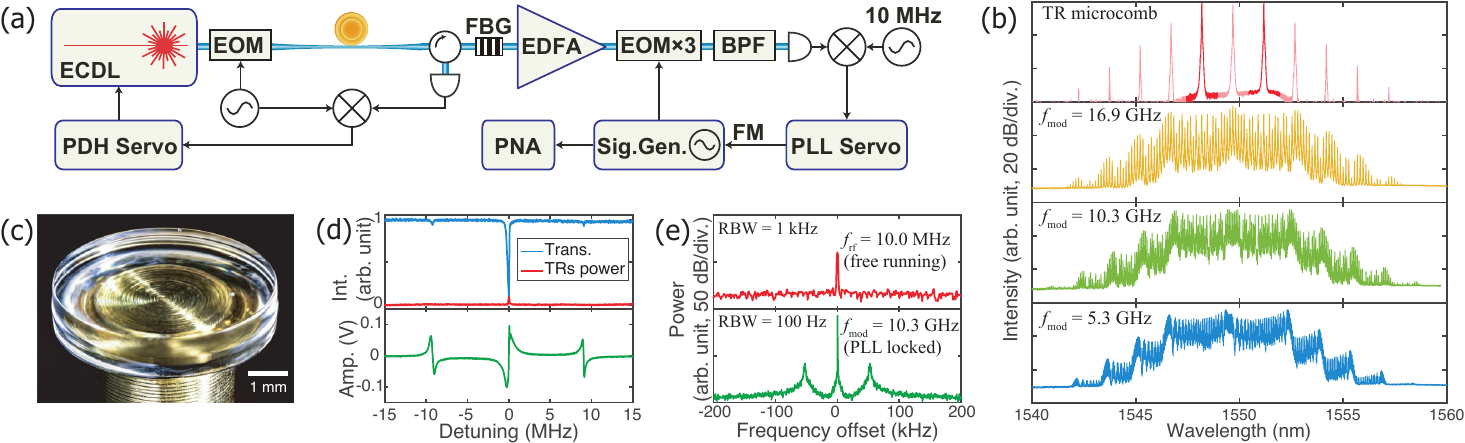} 
\caption{Experimental realization of the TR-EOFD scheme. (a) An ECDL is locked to a microresonator mode with PDH technique to generate a TR microcomb. After amplification using an erbium-doped fiber amplifier (EDFA), the microcomb is phase modulated by three EOMs. An optical bandpass filter (BPF) with a passband centered at 1549.5 nm is used to filter out the overlapped EOCs to generate a beat frequency around 10 MHz. The PLL servo phase-locks $f_\mathrm{mod}$ to the TR oscillation frequency, thus generating low-noise microwaves whose phase noises are measured with a phase noise analyzer (PNA). (b) The optical spectra of the original TR microcomb and the combs after electro-optical modulation at three different modulation frequencies respectively. The two first-order TR sidebands are in deep red. (c) A photo of the MgF$_2$ microresonator. (d) The laser-sweeping spectra of the mode resonance and the generated microcomb power, and the corresponding PDH error signal. The PDH modulation frequency is 9 MHz. (e) Electrical spectra of the free-running $f_\mathrm{rf}$ and the phase-locked $f_\mathrm{mod}$ measured with an electrical spectrum analyzer (ESA).
}\label{fig2}
\end{figure*}

Our in-depth analysis detailed in Appendices A, B and C shows that, as long as the phase noise is not limited by thermal noise, the TR-EOFD scheme could yield low-noise microwaves with spectral purities outperform those of DKS oscillations. In Fig.\,\ref{fig1} (b) we plot the calculated quantum noises resulting from fundamental phase diffusion \cite{matsko2005optical,matsko2013timing,matsko2019hyperparametric} and the simulated technical noises for the TR oscillation (of frequency $f_{\rm TR} = 16 \nu_{\rm FSR}$, where $\nu_{\rm FSR} =$ 12 GHz is the frequency of the microresonator FSR), the DKS oscillation and the synthesized microwave of the frequency of FSR, respectively. The resonator-related parameters used in the calculations and the simulations are similar to those of the crystalline microresonator in our experiments, and the technical noises include both the amplitude-modulation (AM) -to-phase-modulation (PM) noise caused by laser relative intensity noise (RIN) and the PM-to-PM noise induced by laser phase noise (see details of the simulation process and the discussion on the noise transfer mechanism in Appendix B). One should note that changing the absolute values of the laser RIN and phase noise would not change the relative levels of the technical noises as the noise transfer functions do not depend on the input noise amplitude. This result shows that TR-EOFD scheme takes advantage of the frequency division principle and generates microwaves with phase noises that are lower than those of DKS oscillations in the same microresonator by up to 20 dB.

\section{Variable low-noise microwave synthesis}

We carry out experiments using the setup depicted in Fig.\,\ref{fig2} (a). With the Pound-Drever-Hall (PDH) technique a 1550-nm external-cavity diode laser (ECDL) is frequency-locked to a high-$Q$ ($\sim1\times10^9$) resonance in a magnesium fluoride (MgF$_2$) whispering-gallery-mode microresonator. The microresonator is placed in a metal enclosure without active temperature control, only to shield it from ambient air flow. Approximately 4 mW of optical power is coupled into the microresonator via a tapered fiber, generating two first-order TR sidebands with a frequency gap of 0.372 THz. After filtering and amplification, the TR sidebands are modulated with three EOMs. In this proof-of-concept experiment, we employ a commercial microwave signal generator instead of a VCO to provide $f_\mathrm{mod}$. With a 10-MHz reference signal (whose phase noise is below $-130$ dBc$/$Hz at all relevant offset frequencies between 10~Hz and 10~MHz) as the local oscillator, a PLL servo feeds back to the frequency-modulation (FM) port of the microwave signal generator to execute the lock. Figure \ref{fig2} (b) shows the TRs spectrum and the comb spectra that are electro-optically modulated at three different frequencies, corresponding to a division factor of 22, 36, and 70, respectively. As shown in Fig.\,\ref{fig2} (d), the TRs conversion efficiency is below $10\%$. We note that the conversion efficiency is intentionally kept low so the TR oscillation is close to the three-mode hyperparametric oscillation studied in \cite{matsko2005optical,matsko2019hyperparametric}, although it can be raised above $20\%$ without entering the chaotic regime \cite{godey2014stability} by increasing the pump power. Figure \ref{fig2} (e) shows the spectra of $f_\mathrm{rf}$ when the PLL is not activated and the signal of $f_\mathrm{mod} = 10.3$ GHz when the PLL is engaged, respectively. The control bandwidth of approximately 50 kHz is manifested as the servo bumps at the corresponding offset frequency in the PLL-locked spectrum of $f_\mathrm{mod}$.

\begin{figure}
\centering
\includegraphics[width=0.96\columnwidth]{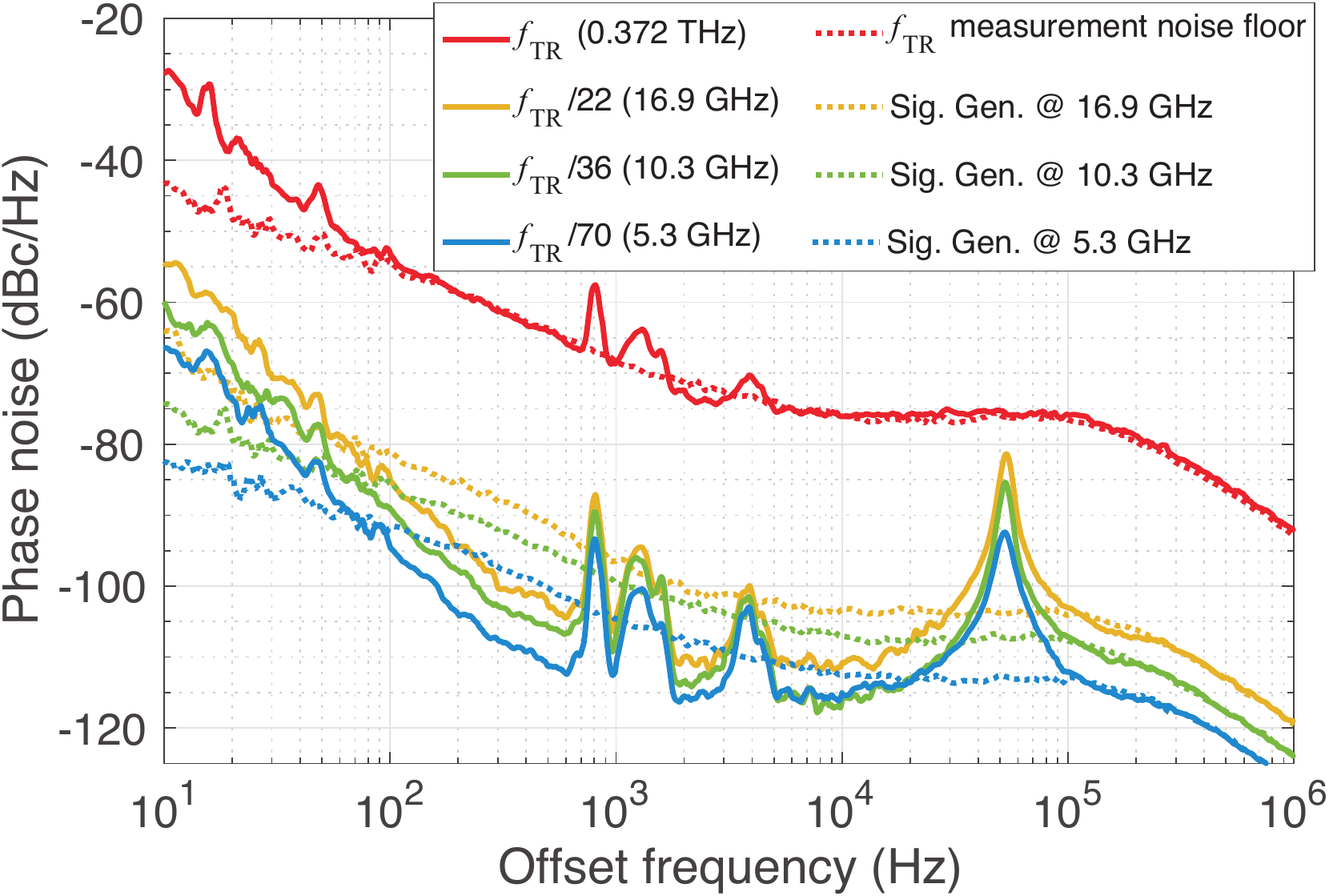} 
\caption{Phase noise spectra of (solid lines) synthesized microwave signals and (dotted lines) the corresponding uncontrolled $f_\mathrm{mod}$. As an indirect measurement of the phase noise of the 0.372-THz TR oscillation, the phase noise spectrum of $f_{\rm rf}$ when the PLL is deactivated is also presented, showing that at frequencies above 100 Hz the measurement is limited by the multiplication noise of the uncontrolled $f_{\rm mod}$. The noise peaks around 1 kHz is caused by the phase noise of the pump laser.
}\label{fig3}
\end{figure}

The phase noise spectra of the synthesized microwaves are presented in Fig.\,\ref{fig3}, along with the phase noise spectrum of the free-running $f_\mathrm{rf}$. As an indirect measurement of the TR oscillation phase noise, the noise spectrum of $f_\mathrm{rf}$ contains the intrinsic noise of the microwave signal generator compounded with a multiplication effect. Comparing with the $f_\mathrm{rf}$ phase noise at frequencies below 100 Hz where the noise level is above the microwave multiplication noise floor, the synthesized microwaves exhibit a phase noise reduction of approximately 27, 31, and 37 dB respectively, showing excellent agreement with the expected noise reduction effect. Out of the PLL control bandwidth, the phase noise amplitudes of the synthesized microwaves are the same as the intrinsic noise levels of the microwave source. Remarkably, the synthesized 10.3-GHz microwave phase noise spectrum reaches $-60$, $-90$, and $-110$ dBc$/$Hz at offset frequencies of 10, 100 and 1000 Hz respectively. Besides demonstrating an improvement by nearly 40 dB over phase noise levels previously achieved with crystalline microresonator TRs \cite{liang2015high,saleh2016phase}, the TR-EOFD scheme is unprecedented in the flexibility of generating low-noise microwaves of different frequencies from a single TR source.

\begin{figure*}[hpt]
\centering
\includegraphics[width=0.99\textwidth]{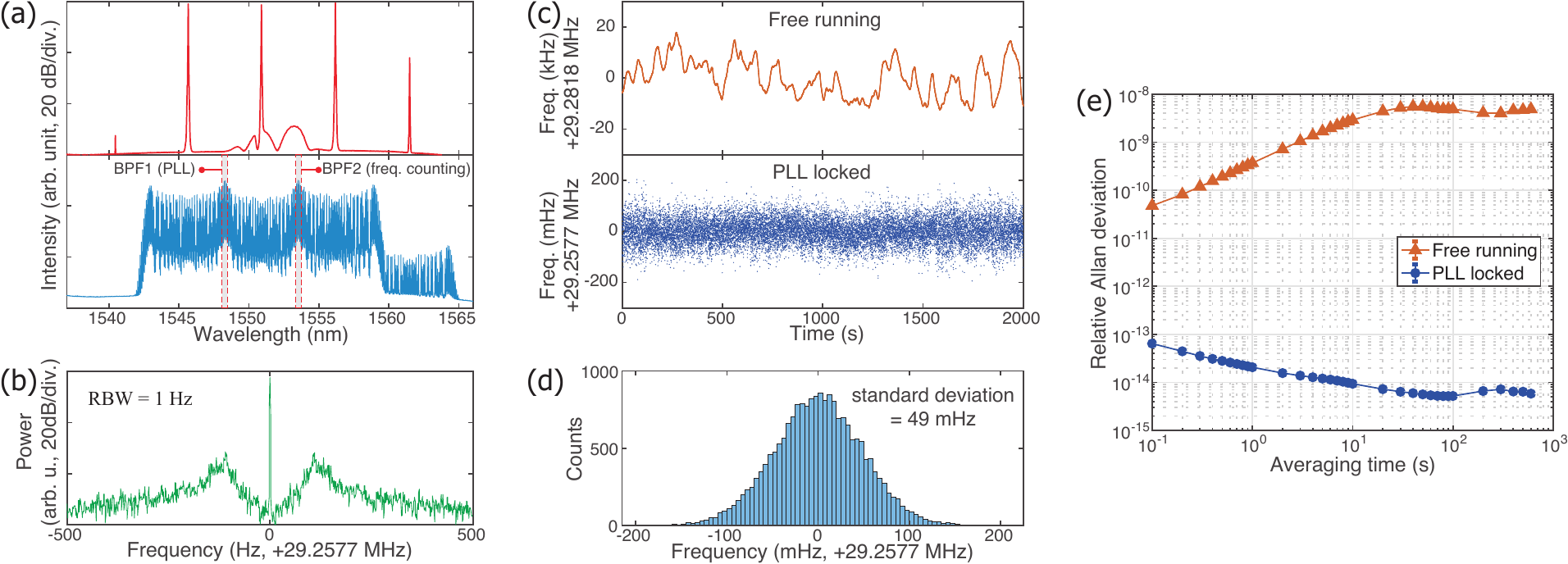} 
\caption{Phase-coherent microwave-to-THz link based on actively frequency-stabilized Turing rolls. (a) The optical spectra of the TRs before (top) and after (bottom) the electro-optic modulation. (b) Spectrum of $f_\mathrm{rf2}$ when the PLL is activated. (c) Measured frequencies of $f_\mathrm{rf2}$ without (top) and with (bottom) the PLL control. (d) The histogram of the counted frequencies with the PLL activated. The standard deviation is indicated. (e) Computed modified Allan deviations of the frequency traces presented in (c). The error bars are too small to be visible.
}\label{fig4}
\end{figure*}

\section{Microwave-to-THz link}

Next, we reverse the path of coherence transfer by stabilizing a THz TR oscillation to a microwave signal. To demonstrate the accessibility of TRs with different oscillation frequencies and the broad spectral span of the EOCs, in a different transverse mode family (i.\,e., modes with dissimilar polar and/or radial field maxima numbers) we generate TR sidebands that are apart from the pump laser by 0.68 THz. Without filtering out the pump line, three intensive sets of EOCs are generated by setting $f_\mathrm{mod}$ to 11.5 GHz (i.\,e., a division factor of 59). We use two optical bandpass filters (BPF1 and BPF2) to select the overlapped edges of the EOCs to generate two rf signals ($f_\mathrm{rf1}$ and $f_\mathrm{rf2}$) that are of the identical frequencies of $\sim29$ MHz. One of the signals ($f_\mathrm{rf1}$) is used to generate the error signals for the PLL, while the other ($f_\mathrm{rf2}$) is for spectral measurement and frequency counting of the THz oscillation (see Fig.\,\ref{fig4} (a)). Instead of controlling $f_\mathrm{mod}$ with the PLL, in this experiment the PLL servo feeds back to the pump power via an acousto-optic modulator (AOM). Mainly relying on the thermal expansion and thermo-optic effects with intracavity power modulation, this approach has a low control bandwidth of a few tens of hertz. Nevertheless, the PLL can be tightly locked, as shown by the spectrum of $f_\mathrm{rf2}$ (see Fig.\,\ref{fig4} (b)) whose linewidth is limited by the ESA's minimum resolution bandwidth (RBW) of 1 Hz. We use a frequency counter with a gate time of 0.1 s to count the frequency of $f_\mathrm{rf2}$ for more than half an hour in free-running and PLL-locked conditions respectively. The counting results are presented in Fig.\,\ref{fig4} (c) and (d), showing that the intracavity power control suppresses the frequency instability of $f_\mathrm{rf2}$ to a sub-Hz level. The computed modified Allan deviations of the fractional frequency instabilities (with a nominal frequency of 0.68 THz) are displayed in Fig.\,\ref{fig4} (e). At averaging times above 10 s the free-running frequency instabilities are reduced by nearly six orders of magnitude by the feedback control, reaching relative Allan deviations below $1\times10^{-14}$. We have verified that the frequency instability attributed to the microwave multiplication is below $1\times10^{-15}$ at all averaging times. Therefore this measurement reliably evaluates the frequency stability of the THz oscillation.


\section{Discussion}

By leveraging the frequency division technique, we unleashed new capabilities of TRs in synthesizing microwaves with high spectral purity. Compared with the previous work of using DKS to purify an injected microwave signal \cite{weng2019spectral}, the TR-EOFD scheme not only generates microwaves with variable frequencies but also has a significantly larger frequency locking range that is only limited by the maximum frequency deviation of the frequency-modulated microwave source. The demonstrated microwave phase noise performance is very close to that obtained with a DKS state in a similar microresonator \cite{lucas2020ultralow}, which, to the best of our knowledge, is to date the lowest phase noise level achieved with a microresonator Kerr oscillator. One should note that, in \cite{lucas2020ultralow} the DKS state was operated at a ``quiet point'' -- a particular pump-resonance detuning position where the noise transduction is minimized due to the mutual cancelation of multiple detuning-to-repetition-rate coupling effects \cite{yi2017single}. Tracking down the quiet point is non-trivial, and its operation is delicate and vulnerable to perturbations in temperature and coupling condition. In contrast, the TR-EOFD scheme offers great ease in its implementation, and the flexibility in switching microwave frequencies provides unprecedented convenience for practical applications.

Additionally, the system is capable of synthesizing ultrastable THz oscillations from rf sources. Compared with prior endeavors in generating low-noise THz radiation with DKS microcombs \cite{zhang2019terahertz,drake2019terahertz,tetsumoto2020300}, the state of TRs explored here yields not only inherently low power threshold and high conversion efficiency but also simplicity in system development. With future effort in packaging and integration, the technology can be fruitfully utilized in advanced THz imaging \cite{mittleman2018twenty} and next-generation communication networks \cite{nagatsuma2016advances}.


\begin{acknowledgments}
This publication was supported by Contract D18AC00032 (DRINQS) from the Defense Advanced Research Projects Agency (DARPA), Defense Sciences Office (DSO) and funding from the Swiss National Science Foundation under grant agreement No.\,192293.
\end{acknowledgments}

\section*{Appendix A: Quantum noises of TR oscillations}
\setcounter{equation}{0}
\renewcommand\theequation{A\arabic{equation}}

The theory \cite{matsko2013timing} for computing quantum noise contribution to the phase noise of DKS repetition rate has been verified both numerically \cite{bao2021quantum} and experimentally \cite{jeong2020ultralow}. In this section we simulate the quantum-noise-limited phase noise spectra of the TR oscillations by numerically integrating the modified LLE that contains the Langevin noise term. The master equation can be written as:
\begin{multline}
\label{lle}
\frac{\partial{A}}{\partial t} - i \frac{1}{2} D_{2} \frac{\partial^2{A}}{\partial \phi^2} - i g {|A|^2 A}  = -\left( {\frac{\kappa}{2} + i(\omega_0 - \omega_\mathrm{p}) } \right){A} \\
+ \sqrt{\frac{\kappa _{\rm ex} P_{\rm in}}{\hbar \omega_0}} + f(t,\phi)
\end{multline}
where $A$ is the slowly varying field amplitude, $\phi$ is the angular coordinate in the co-rotating frame that is related to the round-trip fast time coordinate $\tau$ by $\phi = \tau \times D_1$ (where $D_1 = 2 \pi \nu_{\rm FSR}$), ${\kappa}$ is the cavity decay rate, ${\kappa_{\rm ex}}$ is the external coupling rate, and $P_{\rm in}$ is the pump power. Here $g =\frac{\hbar\omega_{0}^{2}cn_{2}}{n_0^{2}V_{\rm eff}}$ is the single photon induced Kerr frequency shift, where  $n_0$ and $n_{2}$ are the refractive and nonlinear optical indices, $V_{\rm eff}$ is the effective mode volume, and $c$ is the speed of light. The Langevin force term $f(t,\phi)$ obeys \cite{matsko2019hyperparametric}:
\begin{equation}
\label{langevin}
\langle f(t,\phi) f^*(t',\phi') \rangle = \frac{\kappa \hbar \omega_0 \delta(t-t',\phi - \phi')}{\Delta \tau}
\end{equation}
where $\Delta \tau = \frac{\Delta \phi}{D_1}$ is the fast time simulation resolution. In the simulations we set $\Delta \phi = \frac{2 \pi}{256}$.

Using step size of a single cavity round trip, we simulate for consecutively $1\times10^6$ round trips. The phase evolution of the TR oscillation is computed with \cite{matsko2015noise}:
\begin{equation}
\label{TR_phase}
\Psi(t) = {\rm Arg}[e^{i 2 \pi (m \nu_{\rm FSR} + \delta f)t} (a_{-} a_{0}^* + a_{0} a_{+}^*)]
\end{equation}
where $a_0$, $a_{-}$ and $a_{+}$ are the complex intracavity fields of the pumped mode, the low-frequency first-order TR sideband and the high-frequency first-order TR sideband at slow time $t$, respectively. These complex fields can be derived by applying Fourier transformation to the slowly varying field amplitude $A$ in Eq.\,\ref{lle}. Without loss of generality, $\delta f$ is included in the term that represents the TR oscillation frequency for minor frequency deviation from $m \nu_{\rm FSR}$, where $m$ is the number of TRs per round trip. We note that only the light in the pumped mode and the two first-order sidebands are taken into consideration as only they are relevant in our experiments.

\begin{figure}[hbt!]
\centering
\includegraphics[width=0.86\columnwidth]{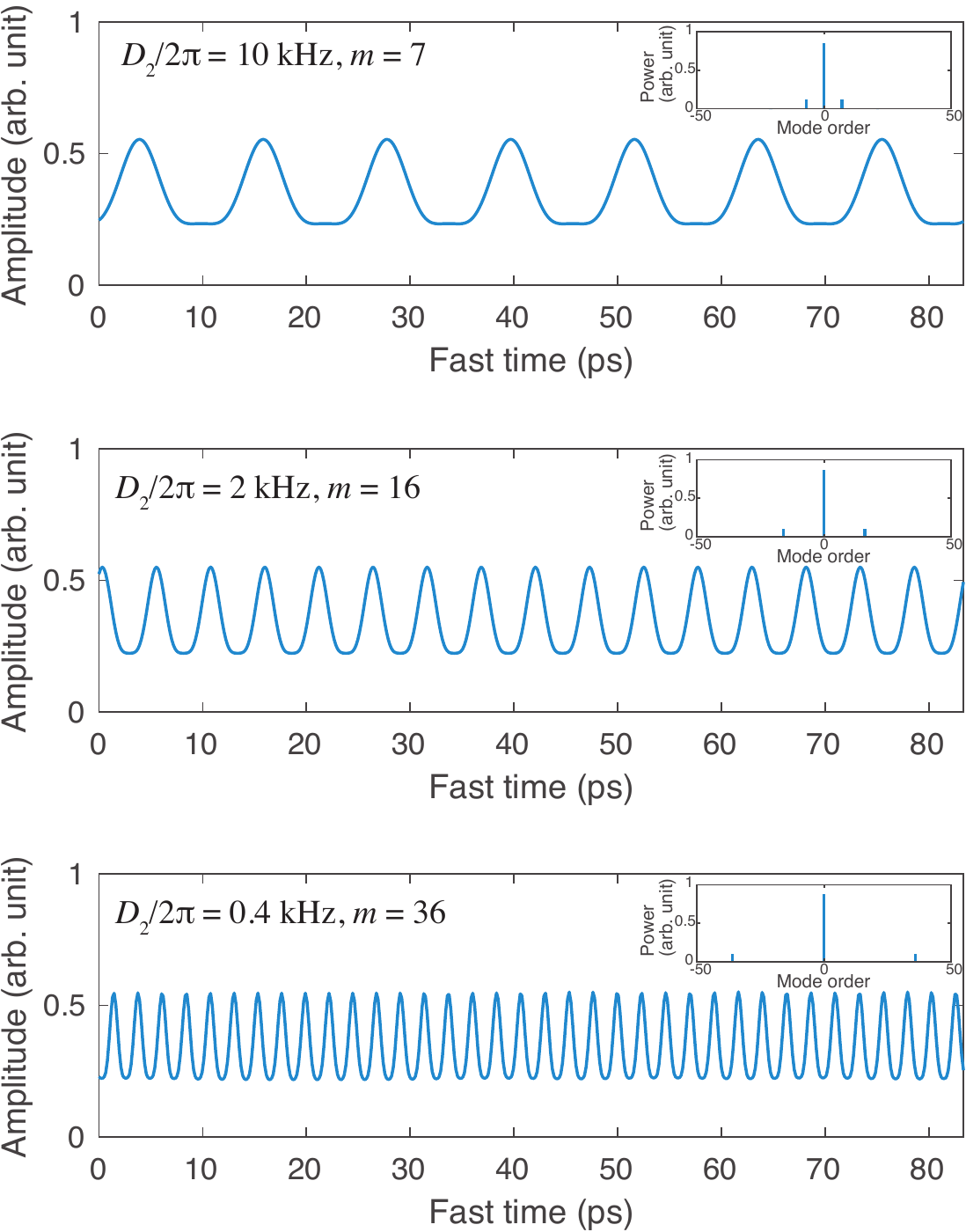} 
\caption{The profiles of the intracavity field amplitude of the simulated TRs with three different $D_2$ settings. The values of $D_2$ and the numbers of TRs are shown in the corresponding panels. The insets plot the relative intracavity powers of the pumping laser and the first-order TR sidebands.
}\label{figS1}
\end{figure}

\begin{figure}[hbt!]
\centering
\includegraphics[width=0.88\columnwidth]{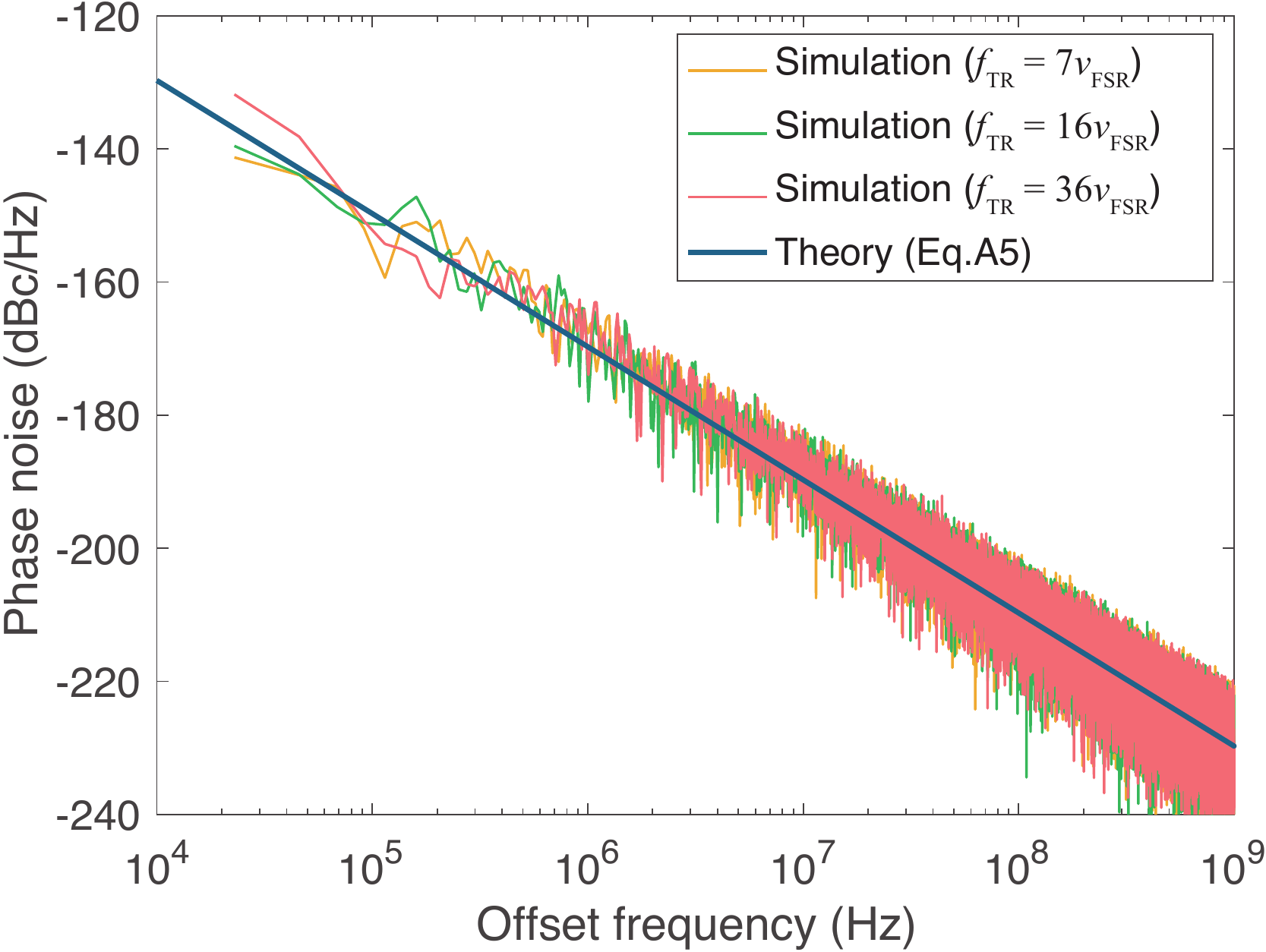} 
\caption{Numerically simulated quantum-noise-limited phase noise spectra for three different TR oscillation frequencies show that the phase noise levels are independent of the roll number $m$. The theoretical estimation is also presented for comparison.
}\label{figS2}
\end{figure}

To numerically test the quantum phase noises of TR oscillations with different roll numbers ($m$), besides using $\frac{D_2}{2\pi} = 2$ kHz that is close to the experimental parameter, we also repeat the simulations with a smaller $D_2$ ($= 2\pi\times0.4$ kHz) and a larger $D_2$ ($= 2\pi\times10$ kHz), respectively. The number of rolls (close to MI threshold)  can be estimated according to \cite{godey2014stability} using:
\begin{equation}
\label{eqs1}
m = \sqrt{\frac{2}{\beta}(\alpha - 2\rho)}
\end{equation}
where $\alpha = \frac{2 (\omega_0 - \omega_\mathrm{p})}{\kappa}$, $\beta = -\frac{4 \pi D_2}{\kappa}$, and $\rho$ is calculated to be 1.3 with $F^2 = (1+(\rho-\alpha)^2)\rho$ and $F = \sqrt\frac{8 g \kappa_{\rm ex} P_{\rm in}}{\kappa^3 \hbar \omega_0}$. We estimate $m$ to be 7.23, 16.17, and 36.15 for $\frac{D_2}{2\pi} =$ 10, 2, and 0.4 kHz, respectively, showing excellent agreement with the simulation results in Fig.\,\ref{figS1}. One should note that, in practice the number of rolls might be tunable by temperature tuning via the mode-interaction-induced avoided mode crossing phenomenon \cite{weng2015mode,xue2015normal}. 

In Fig.\,\ref{figS2} we plot the computed phase noise spectra of the three TR oscillation frequencies of $f_\mathrm{TR} =$ 7$\nu_\mathrm{FSR}$, 16$\nu_\mathrm{FSR}$, and 36$\nu_\mathrm{FSR}$, respectively. One can see that these phase noise spectra have the same magnitude, despite the fact that the carrier frequencies of the TR oscillations are quite different. This confirms that the quantum noise of a TR oscillation does not depend on $m$, which agrees with the theoretical derivation in \cite{matsko2019hyperparametric}, which is written as:
\begin{equation}
\label{eqs2}
\mathcal{L}_{\rm QN}(f) = \frac{\kappa^2 \hbar \omega_0}{4 \pi^2 f^2 P_\mathrm{out}}
\end{equation}
where $P_\mathrm{out}$ is the output power of the first-order sideband for which we use the simulation result to obtain. The theoretically calculated quantum phase noise is also presented in Fig.\,\ref{figS2}, showing excellent agreement with numerical simulations.

From the analysis presented in this section, one can conclude that, as long as the TR oscillation frequency is within the frequency span of the EO comb, the TR-EOFD scheme benefits from a large roll number $m$ in quantum-noise-limited phase noise performance.

\section*{Appendix B: Technical noises of TR oscillations}
\setcounter{equation}{0}
\renewcommand\theequation{B\arabic{equation}}

Most of the time technical noises are the dominant noises that degrade Kerr oscillation spectral purity, especially at offset frequencies below 100\,kHz. Technical noises can be either of thermal nature or non-thermal. For thermal noises such as resonator temperature fluctuations caused by ambient temperature change and intracavity power absorption due to laser RIN or the instability of coupling condition, we will briefly discuss their impact together with the influence of fundamental thermal noise in the next section. In this section we focus our attention on non-thermal technical noises including laser-RIN-induced AM-to-PM noise and laser-phase-noise-induced PM-to-PM noise. The noise conversion is simulated using methods similar to \cite{matsko2015noise}. To simulate the technical phase noise spectrum of the DKS repetition rate presented in Fig.\,\ref{fig1} (b), the pump power is set to 20\,mW and the pump-resonance detuning is $-10 \kappa$. All the other parameter values are the same as those used for TR oscillation simulations. Like the method in \cite{matsko2015noise}, a small third-order dispersion $\frac{D_3}{2\pi} = -4$ kHz is included in the simulations in order to effectively induce the noise transfer mechanism. (In principle, even if the resonator has no higher-order dispersion, $D_2$ is sufficient to cause noise transfer. However, in reality the higher-order dispersion is always present, and the noise transfer caused by the higher-order dispersion is significantly stronger.)

\begin{figure}[hbt!]
\centering
\includegraphics[width=0.92\columnwidth]{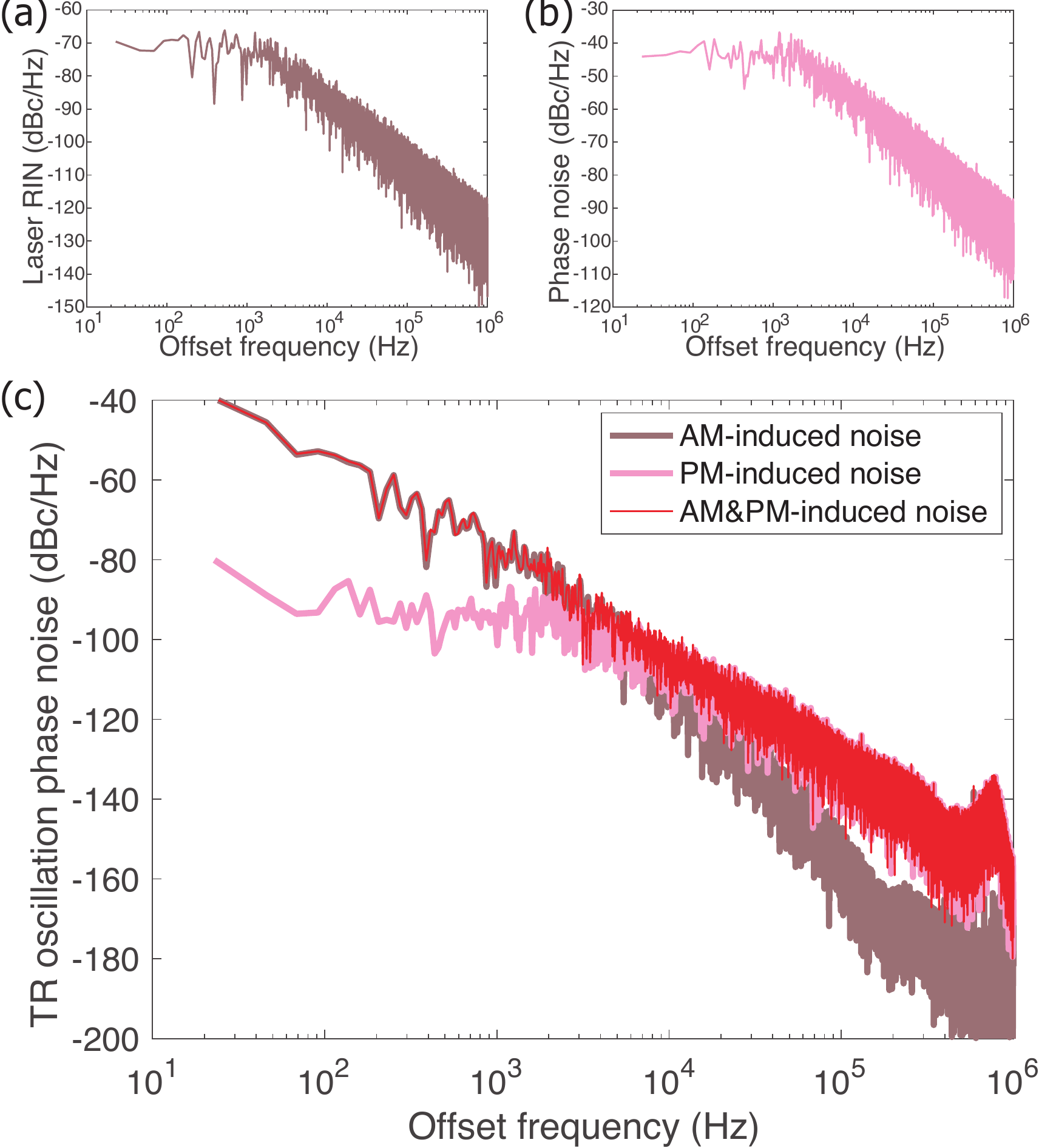} 
\caption{(a) Spectrum of the laser RIN that is used for technical noise simulation. (b) Spectrum of the computer-generated laser phase noise that is for numerical simulation. (c) Phase noise spectra of the TR oscillation simulated with input noise of (brown) laser RIN only, (pink) laser phase noise only, and (red) both laser RIN and phase noise.
}\label{figS3}
\end{figure}

Figure \ref{figS3} shows the simulated technical phase noises of TR oscillations. The input noises, including the laser RIN and the optical phase noise, are synthesized to possess a power spectral density slope of $f^{-2}$ (see Fig.\,\ref{figS3} (a) and (b)). This slope of the noise spectra is consistent with most realistic laser noises. We confirm that the TR oscillation phase noises converted from laser RIN and phase noises are independent and that the TR oscillation phase noise resulted from compound noises is equal to the sum of RIN-only-induced noise and laser-phase-noise-only-induced noise (see Fig.\,\ref{figS3} (c)). In addition, the different slopes exhibited by the spectra of RIN- and laser-phase-noise-induced phase noises are in agreement with the characteristics of the noise transfer functions revealed in \cite{matsko2015noise}.

\begin{figure}[hbt!]
\centering
\includegraphics[width=0.86\columnwidth]{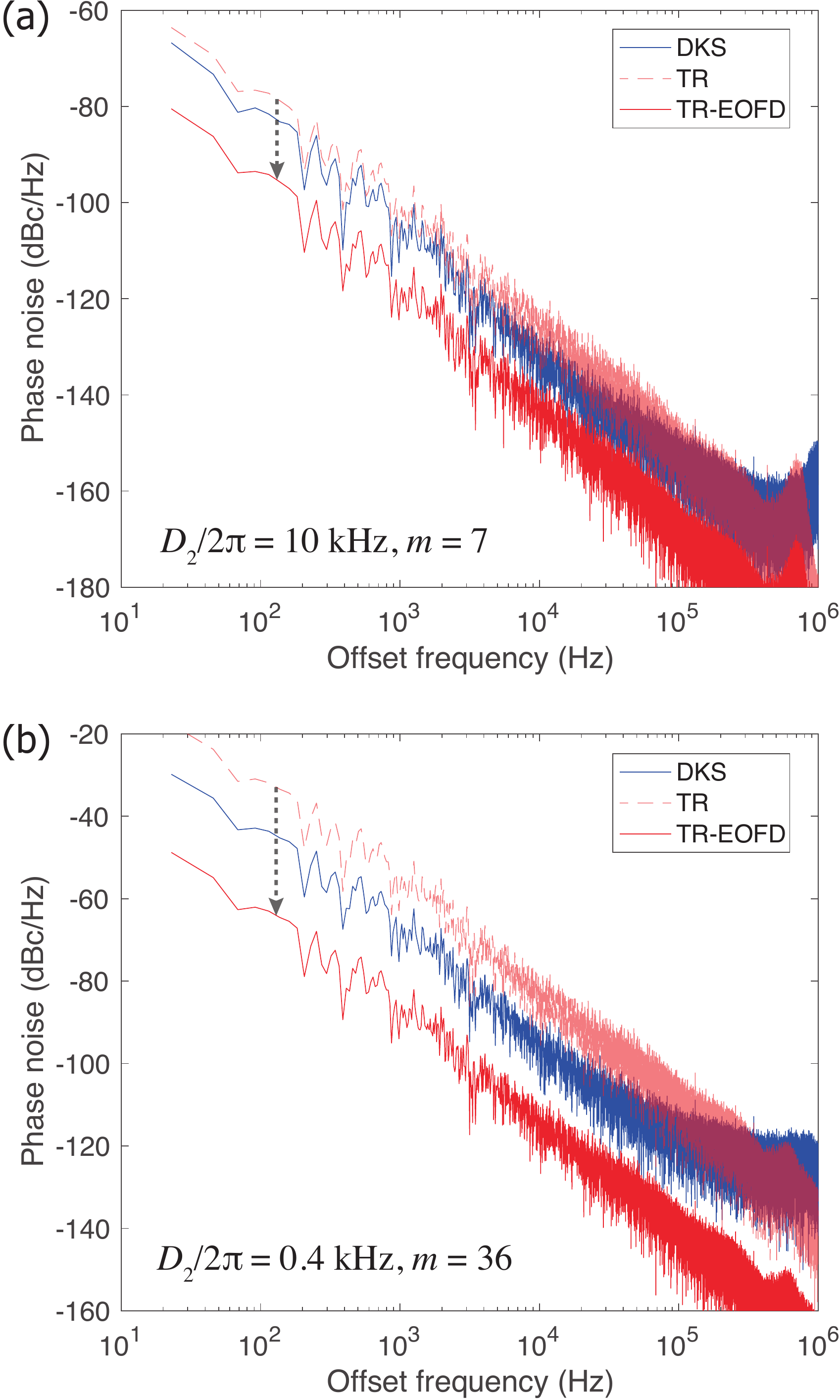} 
\caption{Simulated technical phase noise spectra of the DKS oscillation, the TR oscillation, and the TR-EOFD-synthesized microwave with the same frequency of the DKS oscillation when $\frac{D_2}{2\pi}$ is set to 10 kHz (a) and 0.4 kHz (b) respectively.
}\label{figS4}
\end{figure}

We also repeat the same noise transfer simulation with different $\frac{D_2}{2\pi}$ values of 10 and 0.4 kHz respectively while $D_3$ is kept the same. Figure \ref{figS4} presents the simulation results. While the phase noise amplitudes differ with different $D_2$ because of the different recoil strengths that are related to the change in microcomb spectral span, the relative levels of the phase noises are very similar. Just like the result shown in Fig.\,\ref{fig1} (b), the microwaves yielded by the TR-EOFD scheme exhibit phase noise amplitudes that are lower than those of DKS oscillations by 10 to 20 dB. Physically, the smaller technical noise transfer magnitude of TR state might be related to its narrower optical spectral span. Under the influence of the same higher order dispersion effect, a wider optical spectral span usually results in larger spectral asymmetry, thus leading to larger spectral recoil \cite{lucas2017detuning} and more prominent technical noise transfer. Since the optical spectral span of DKS is normally much wider than that of the TRs, the technical noise transfer in the repetition rate is also severer for the DKS state.

\section*{Appendix C: Discussion on thermal noises}
\setcounter{equation}{0}
\renewcommand\theequation{C\arabic{equation}}

Since a Kerr oscillator's fractional frequency change caused by the change of the resonator temperature, either fundamental or technical, is independent of the nominal frequency of the Kerr oscillation, in principle the TR-EOFD scheme would not make any phase noise improvement if the Kerr oscillation (either TRs or DKS) spectral purity is thermal-noise-limited. Yet, from a practical point of view, the TR-EOFD scheme may still be preferable. Because the TR state is excited at near-threshold power level, the coupled-in laser power is relatively low (i.\,e., below a few milliwatts for most high- or ultrahigh-$Q$ resonators). On the contrary, in order to generate stable DKS state and to have soliton steps that are wide enough, the pump power is often set to be higher than the MI threshold by one to two orders of magnitude. Consequently, the usage of higher power could result in larger temperature fluctuations with the same laser RIN amplitude.

Thermo-refractive noise often dictates the fundamental resonance frequency noise limit in dielectric microresonators \cite{matsko2007whispering} at relatively low Fourier frequencies. It was indicated in \cite{yang2021dispersive} that for DKS in on-chip SiO$_2$ microresonators the thermo-refractive noise is higher than the quantum noise in the DKS oscillation fundamental phase noise at offset frequencies below a few kHz. While the finite element method is often used to obtain a more accurate estimation of the fundamental thermo-refractive noise \cite{weng2018ultra,huang2019thermorefractive}, here we use the analytical expression derived in \cite{matsko2007whispering} to compute the thermo-refractive-noise-limited phase noise of the TR oscillation in the resonator used in our experiment, and we compare it with the quantum noise. The formula for computing the thermo-refractive-noise-limited phase noise is:
\begin{widetext}
\begin{equation}
\label{TRN}
\mathcal{L}_{\rm TRN}(f) =\frac{1}{2f^2} (\nu_{\rm FSR} \frac{1}{n_0} \frac{dn}{dT})^2 \frac{k_{\rm B} T^2}{\rho C_{\rm p} V_{\rm eff}} \frac{R^2}{12 D} \left[1 + (\frac{R^2}{D} \frac{2 \pi f}{9 \sqrt{3}})^{3/2} + \frac{1}{6}(\frac{R^2}{D}\frac{2 \pi f}{8 l^{1/3}})^2 \right]^{-1}
\end{equation}
\end{widetext}
where $k_{\rm B}$ is the Boltzmann's constant, $\rho$ is the density of the dielectric material, $C_{\rm p}$ is the specific heat capacity of the material, and the thermal diffusion coefficient $D$ is related to the thermal conductivity $K$ as $D = \frac{K}{\rho C_{\rm p}}$. The whispering gallery mode azimuthal number $l$ can be estimated by $l \approx \frac{R n_0 \omega_0}{C}$. The thermo-optic coefficient ($\frac{dn}{dT}$) of MgF$_2$ for transverse electric (TE) mode (o-ray in z-cut disks) is $8.4\times10^{-7}$K$^{-1}$ \cite{ghosh1998handbook}.

\begin{figure}[hbt]
\centering
\includegraphics[width=0.86\columnwidth]{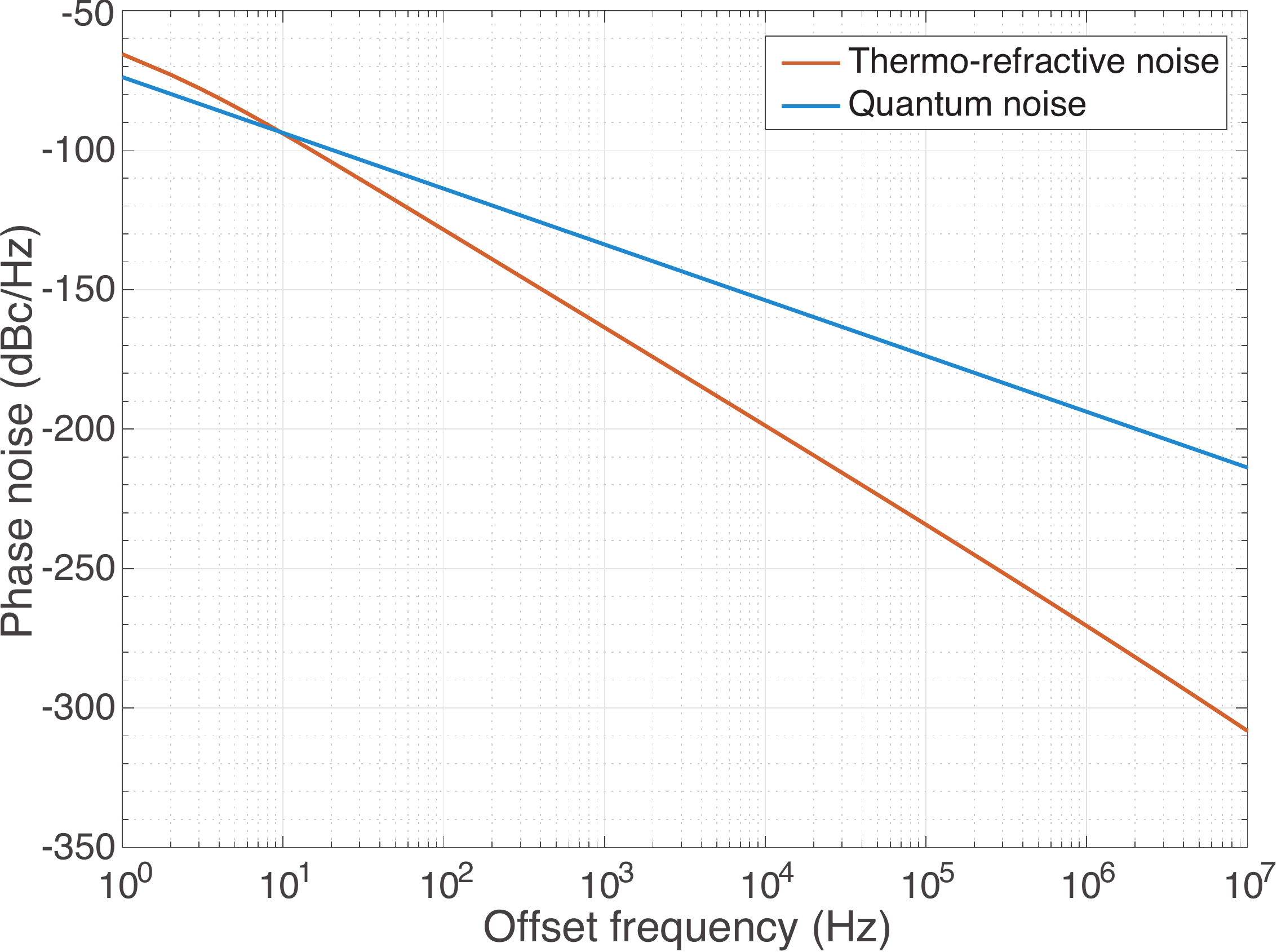} 
\caption{Computed fundamental phase noises including the thermo-refractive noise and the quantum noise of a microwave that is synthesized with the TR-EOFD scheme. The TR oscillation is of a roll number of 16. The frequency of the synthesized microwave is equal to $\nu_{\rm FSR}$.
}\label{figS5}
\end{figure}

In Fig.\,\ref{figS5} we plot the thermo-refractive noise limit and the quantum noise limit for the phase noise of the microwave frequency that is obtained with EOFD from a TR oscillation with a roll number of 16. This comparison shows that the thermo-refractive noise dominates the fundamental noise floor only at offset frequencies below 10 Hz, which means that practically the thermo-refractive noise is unlikely to limit the phase noise performance since at low frequencies technical noises would set the noise level. It is only for microresonators with much smaller effective mode volumes and larger thermo-optic coefficients (such as those on-chip SiO$_2$ and Si$_3$N$_4$ microresonators in \cite{yang2021dispersive,huang2019thermorefractive}) that thermo-refractive noise might be the dominant fundamental phase noise type in certain frequency ranges.



%

\end{document}